# Experimental Realization of Full Control of Reflected Waves with Subwavelength Acoustic Metasurfaces

Yong Li[1,§], Xue Jiang[1], Rui-qi Li[1], Bin Liang[1,2,*], Xin-ye Zou[1], Lei-lei Yin[2] and Jian-chun Cheng[1*]

[1]Key Laboratory of Modern Acoustics, MOE, Department of Physics, Collaborative Innovation Center for Advanced Microstructures, Nanjing University, Nanjing 210093, China

[2]Imaging Technology Group, Beckman Institute, University of Illinois at Urbana-Champaign, Urbana, Illinois 61801, USA

[§]Current address: Department of Physics, Hong Kong University of Science and Technology, Clear Water Bay, Kowloon, Hong Kong, China

∗ To whom correspondence should be addressed:  B. L. (liangbin@nju.edu.cn) or J. C. C. (jccheng@nju.edu.cn).

# ABSTRACT

Metasurfaces with subwavelength thickness have exhibited unconventional phenomena in ways that could not be mimicked by traditional materials. Here we report the analytical design and experimental realizations of acoustic metasurface with hitherto inaccessible functionality of manipulating the reflected waves arbitrarily. By suitably designing the phase shift profile with $2\pi$ span induced by labyrinthine units, the metasurface can reflect acoustic waves in an unusual yet controllable manner. Anomalous reflection and ultrathin planar lens were both demonstrated with carefully designed metasurfaces. Remarkably, the free manipulation of phase shifts offers great flexibility in the design of non-paraxial acoustic self-accelerating beams with arbitrary trajectories. With the extraordinary wave-steering ability, the metasurface should open exciting possibilities for designing compact acoustic components with versatile potential and may find a variety of applications ranging from ultrasound imaging to caustic engineering where design of the shape of a focused trajectory of sound is required.

# I. INTRODUCTION

The past few years have witnessed a rapid expansion of the field of metasurfaces that are optically thin metal gratings with subwavelength periodicity but able to exhibit strong influence on incident light to generate diverse functionalities such as bending light abnormally, manipulating polarization, and creating ultrathin planar lenses [1-7]. Optical metasurfaces with unusual properties that cannot be duplicated by using traditional optical materials or devices have shown great promises in fully manipulating electromagnetic waves. Acoustic wave is another important form of classical wave, and molding the flow of reflected acoustic waves is of paramount significance to the current research efforts in a large variety of applications ranging from room acoustics to medical ultrasound imaging. For the purpose of arbitrarily tailoring the acoustic propagation, however, full control of the phase of acoustical field is required, which will be hard to accomplish with existing acoustic materials or devices due to limitation in their acoustical properties [8]. In this context, it is expectable that the acoustical metasurface with the ability to control the reflected wave at will, if realized, will be highly intriguing and may lead to revolutionary innovations in the acoustic field and, furthermore, the acoustic-based applications. The recent emergence of acoustic metamaterials [9-19] with properties unavailable in nature have broadened the horizon of acoustic waves and enabled fascinating phenomena such as sub-diffraction imaging [20-22], one-way transportation [23-27], cloaking [28-30], and super absorption [31, 32], etc. So far, however, the researches on acoustic metasurface are scarce [33, 34] and there still remain considerable practical challenges to experimentally realize the full manipulations of acoustic reflected waves with subwavelength metasurfaces.

In this work, we have analytically designed and experimentally demonstrated acoustic metasurfaces capable of arbitrarily manipulating the reflected waves in direction and shape of wavefront. A general scheme of designing an acoustic metasurface is developed on the basis of some kinds of labyrinthine units. It is worth stressing that the proposed scheme provides substantial conceptual advance over the first theoretical frame presented by us with some seemingly-similar labyrinthine units for which the physical mechanism is complicated and cannot be understood easily due to the extra straight air channels [33]. Furthermore, the proposed scheme takes advantage in device performance since the original design needs a great many of zigzag channels to realize the desired phase shifts, which inevitably leads to difficulties in experiments and remarkable viscosity [9]. Thank to such a different design idea, the resulting metasurface can reflect acoustic waves in an unusual yet controllable manner, for which both the underlying physics and the device performance can be well understood with the analytical method developed here. Despite vanishing thickness in comparison to the incident wavelength, acoustic metasurfaces are shown both theoretically and experimentally to be having the distinct properties of realizing abnormal reflection governed by the generalized Snell's law and ultrathin planar lens with desired focal point. Furthermore, we demonstrate the remarkable functionality of generating non-paraxial accelerating beams along arbitrary trajectory, which reveals the great flexibility in the design of acoustic beams enabled by the free manipulation of propagating phase with metasurface. Capable of engineering acoustic wavefront with subwavelength thickness, the acoustic metasurface should open exciting new possibilities for designing versatile compact acoustic components and may apply to important fields ranging from ultrasound imaging to field caustic engineering where design of the shape of a focused trajectory of sound is required.

## II. METHOD AND DESIGN
### A. Analytical reflected filed of the metasurface

Figure 1(a) illustrates the presented acoustic metasurface, which could be characterized by an acoustic normalized admittance $\tilde{\beta}(y) = \rho_0 c_0 / Z_s$. Here $\rho_0 c_0$ is the acoustic impedance of air with $\rho_0$ and $c_0$ referring to the density and velocity of air respectively, and $Z_s(y)$ is acoustic impedance of the metasurface. The admittance considered in our case is a pure acoustic susceptance, which is capable of effectively modulating the phase the reflected waves. The relationship between the susceptance and the phase shift can be expressed as $\tilde{\beta}(y) = i\tan[\phi(y)/2]$ with $\phi(y)$ referring to the phase shift provided by metasurfaces. In the following, frequency domain is considered and the harmonic factor $e^{-i\omega t}$ with $i = \sqrt{-1}$ is omitted. Considering the two-dimensional case with a point source located at $(x', y')$, the Green's function of the free space can be expressed as [8]

$$G(x, y; x', y') = \frac{i}{4} H_0^{(1)}(k_0 R), \qquad (1)$$

where $R$ is distance between the field point $(x, y)$ and the source point $(x', y')$ $R \equiv \sqrt{(x-x')^2 + (y-y')^2}$, $H_0^{(1)}(*)$ is the first kind of Hankel function of the zero order and $k_0 = \omega/c_0$ refers to the wavenumber in air with $\omega$ being the angular frequency. Assuming the metasurface of length $2L$ locates at $x = 0$ and extends from $y = -L$ to $y = L$, the total filed can be expresses as [8]

$$\begin{aligned} p(x,y) = &\int_V G(x, y; x', y') \Im(x', y') dV' \\ &+ \int_{S+D} \left[ G(x, y; x', y') \frac{\partial p(x', y')}{\partial n'} - p(x', y') \frac{\partial G(x, y; x', y')}{\partial n'} \right] dS' \end{aligned} \qquad (2)$$

where $\Im(x, y)$ is the bulk distribution of acoustic sources, which can be simulated by a Dirac function $\Im(x, y) = Q\delta(x - x')\delta(y - y')$ with a normalized strength $Q = 1$, and $\partial/\partial n'$ is the normal derivative of the integral paths S and D, as shown in Fig. 1(a). By substituting Eq. (1) into the Eq. (2), the reflected waves caused by the metasurface can be obtained analytically, as follows

$$p_r(x, y) = -\frac{k_0}{4} \int_{-L}^{L} \tilde{\beta}(y') p(0, y') H_0^{(1)}(k_0 R) \big|_{x'=0} dy' \\ + \frac{i}{4} \int_{-L}^{L} p(0, y') \frac{\partial H_0^{(1)}(k_0 R)}{\partial x'} \bigg|_{x'=0} dy', \tag{3}$$

where $p(0, y)$ is the pressure profile at the metasurface boundary, which can be expressed as

$$\frac{p(0, y)}{2} = \frac{i}{4} H_0^{(1)}(k_0 R) \big|_{x=0} - \frac{k_0}{4} \int_{-L}^{L} \tilde{\beta}(y') p(0, y') H_0^{(1)}(k_0 R) \big|_{x, x'=0} dy' \\ + \frac{i}{4} \int_{-L}^{L} p(0, y') \frac{\partial H_0^{(1)}(k_0 R)}{\partial x'} \bigg|_{x, x'=0} dy'. \tag{4}$$

Equation (4) is a Fredholm integral equation of the second kind, and can be solved through discretization. From Eq. (2), it is found that reflected waves caused by the metasurface can be tailored efficiently by appropriately choosing the expression of $\tilde{\beta}(y)$ or $\phi(y)$. Remarkably, the reflected field can be arbitrarily designed exactly as desired, due to the fact that no assumptions such as far field approximation or Born approximation has been employed in the derivation process. It is also noteworthy that the metasurface with a length of 2$L$ is employed in the derivation and the effect of finite length has already been taken into consideration.

**B. Reflection coefficient of labyrinthine unit**

As shown in Fig. 1(b), the designed metasurface consists of eight different labyrinthine units, in which several stiff plates are arranged in a way to form a zigzag channel [16, 17, 19, 35], which will be demonstrated, in what follows, to be sufficient for constructing the reflected field at will. For acoustic metasurfaces, the phase of the reflected waves by each unit is crucial for the acoustic phenomenon. It is therefore essential to develop a method to obtain the reflected phase to benefit the design of metasurfaces. Here, we developed a method to analytically solve the reflected waves by labyrinthine unit. A specific labyrinthine unit is considered, as shown in Fig. 1(c), which is divided into two kinds of short pipes: the first kind of pipes are labeled in turn as A, C, E, G, I, and the second kind of pipes are labeled in turn as B, D, F, H, J, respectively. The alternate arrangement of these two kinds of pipes can be equivalently regarded as an abrupt variation in the cross section of the pipes. Let $(p_L, u_L)$ denote the sound pressure and the axial components of the velocities, respectively. Here the subscript $L=i$ and $r$ refer to the incident and the reflected fields respectively, and $L=A \sim J$ indicate the fields inside the corresponding channels. The velocity components $u$ are associated with the corresponding sound pressure $p$ via the relationship $ik_0 \rho_0 c_0 u = \partial p / \partial x$.

In acoustics, waves can propagates within subwavelength waveguide in the absence of cutoff frequency. Since the length in $y$ direction is two orders of magnitudes smaller than the wavelength, the field $(p_L, u_L)$ in the regions A (viz., $-w \leq x \leq 0$ and $t \leq y \leq t+d$), B (viz., $-w-d \leq x \leq w$ and $t \leq y \leq t+d+l$), and C (viz., $-2w-d \leq x \leq -w-d$ and $t+l \leq y \leq t+l+d$) can be expanded with normal modes and can be expressed as [8]

$$p_A(x,y) = A^+ e^{ik_0(x+w)} + A^- e^{-ik_0 x}, \tag{5}$$

$$u_A(x,y) = \frac{1}{\rho_0 c_0}\left[A^+ e^{ik_0(x+w)} - A^- e^{-ik_0 x}\right], \tag{6}$$

$$p_B(x,y) = \sum_{n=0}^{\infty} \phi_n(y)\left[B_n^+ e^{ik_{xn}(x+w+d)} + B_n^- e^{-ik_{xn}(x+w)}\right], \tag{7}$$

$$u_B(x,y) = \sum_{n=0}^{\infty} \phi_n(y) \frac{k_{xn}}{k_0 \rho_0 c_0}\left[B_n^+ e^{ik_{xn}(x+w+d)} - B_n^- e^{-ik_{xn}(x+w)}\right], \tag{8}$$

$$p_C(x,y) = C^+ e^{ik_0(x+2w+d)} + C^- e^{-ik_0(x+w+d)}, \tag{9}$$

$$u_C(x,y) = \frac{1}{\rho_0 c_0}\left[C^+ e^{ik_0(x+2w+d)} - C^- e^{-ik_0(x+w+d)}\right], \tag{10}$$

where $\phi_n(y)$ is the transverse eigenmode

$$\phi_n(y) = \sqrt{2-\delta_{0n}} \cos[k_{yn}(y-t)], \tag{11}$$

Here the wavenumber along $y$ direction can be expressed as $k_{yn} = n\pi/(l+d)$ ( $n = 0,1,2,....$ ) and $k_{xn} = \sqrt{k_0^2 - k_{yn}^2}$. The symbols $A^+$, $A^-$, $C^+$ and $C^-$ denote the coefficients in the thin channel, and the $B_n^+$ and $B_n^-$ denote the coefficients of the $n$-th mode with the normal wavenumber $k_n$. Note that the geometry size along $y$ direction of regions represented as A and C (viz. $d$) is much small comparing to the working wavelength. It is therefore possible in this case to simplify the analytical derivation by considering the plane wave component only, i.e., the summations over $n$ have not applied in Eqs. (5), (6), (9) and (10). To obtain more accurate results, for regions represented as B whose geometrical size (viz. $l+d$) is comparable to $a_y$, the summation of normal modes is employed.

Considering that the continuity of pressure should be satisfied on the boundary between the region A and region B at $x=-w$ and the boundary between regions B and C at $x=-w-d$, a transfer matrix can be constructed to connect the coefficients between regions A and C, as follows

$$\begin{bmatrix} C^+ \\ C^- \end{bmatrix} = M_1 \begin{bmatrix} A^+ \\ A^- \end{bmatrix} \tag{12}$$

where

$$M_1 = \begin{bmatrix} \dfrac{\beta^2 - \alpha^2 + 1}{2\beta} & -\dfrac{\beta^2 - (\alpha+1)^2}{2\beta} e^{-ik_0 w} \\ \dfrac{\beta^2 - (\alpha-1)^2}{2\beta} e^{ik_0 w} & -\dfrac{\beta^2 - \alpha^2 + 1}{2\beta} \end{bmatrix} \tag{13}$$

where

$$\begin{aligned} &= \sum_{n=0} \frac{k_0 d}{k_{xn}(l+d)} \times \frac{1+e^{j2k_{xn}d}}{1\ e^{j2k_{xn}d}} \quad \Phi_n^1 \Phi_n^1 \\ &= \sum_{n=0} \frac{k_0 d}{k_{xn}(l+d)} \times \frac{2e^{jk_{xn}d}}{1\ e^{j2k_{xn}d}} \quad \Phi_n^1 \Phi_n^2 \end{aligned} \tag{14}$$

with

$$\Phi_n^1 = \frac{1}{d}\int_t^{t+d} \phi_n(y)dy; \quad \Phi_n^2 = \frac{1}{d}\int_{t+l}^{t+l+d} \phi_n(y)dy. \tag{15}$$

It can be found that the transfer matrix from regions C to E is the same as the above matrix. Therefore, the coefficients in region I can be expressed as

$$\begin{bmatrix} I^+ \\ I^- \end{bmatrix} = M_1^5 \begin{bmatrix} A^+ \\ A^- \end{bmatrix}. \tag{16}$$

Considering the continuity of pressure and volume velocity at the interface between regions I and J, and the fact that the left boundary of region J is hard boundary, the coefficients in region I can be expressed as

$$\frac{I^+}{I^-} = -\frac{1+\alpha}{1-\alpha} \cdot e^{ik_0 w} \tag{17}$$

The pressure filed in the incident region (viz. $x \geq 0$ and $0 \leq y \leq a_y$) can be expressed as

$$p(x,y) = p_i + p_r = e^{ik_0 x} + \sum_{n=0}^{\infty} R_n \psi_n(y) e^{-ik'_{xn} x}, \tag{18}$$

where $\psi_n(r)$ is the transverse eigenmode

$$\psi_n(y) = \sqrt{2-\delta_{0n}} \cos(k'_{yn} y) \tag{19}$$

with the wavenumber along $y$ direction given as $k'_{yn} = n\pi/(l+d+2t)$ and $k'_{xn} = \sqrt{k^2 - k'^2_{yn}}$. Following the same procedure, the transfer matrix connecting the incident region and region A is

$$\begin{bmatrix} A^+ \\ A^- \end{bmatrix} = M_2 \begin{bmatrix} 1 \\ R_0 \end{bmatrix} \tag{20}$$

with

$$M_2 = \begin{bmatrix} \left(1-\dfrac{\beta'}{2}\right)e^{-ik_0 w} & \dfrac{\beta'}{2}e^{-ik_0 w} \\ 1+\dfrac{\alpha'}{2} & -\dfrac{\alpha'}{2} \end{bmatrix} \quad (21)$$

with

$$\alpha' = \dfrac{a_y}{d} - \sum_{n=0}^{\infty}\dfrac{k_0}{k'_{xn}}(\Psi'_n)^2; \quad \beta' = \dfrac{a_y}{d} + \sum_{n=0}^{\infty}\dfrac{k_0}{k'_{xn}}(\Psi'_n)^2 \quad (22)$$

Combining Eqs. (16), (17) and (20), the total transfer matrix can be expressed as

$$R_0 = -\dfrac{(1-\alpha)m_{11} + (1+\alpha)e^{ik_0 w}m_{21}}{(1-\alpha)m_{12} + (1+\alpha)e^{ik_0 w}m_{22}} \quad (23)$$

with

$$M = M_1^5 M_2 = \begin{bmatrix} m_{11} & m_{12} \\ m_{21} & m_{22} \end{bmatrix}. \quad (24)$$

### C. Realization of the full control of reflected waves

Once the complex reflection coefficient is obtained, the impedance of the presented metasurface can be expressed as

$$Z_s = \dfrac{1-R_0}{1+R_0}. \quad (25)$$

Then the admittance of each labyrinthine unit could be expressed as $\tilde{\beta} = \rho_0 c_0 / Z_s = i\tan(\phi/2)$ with $\phi = \arg(R_0)$ being the phase shifts induced by each labyrinthine unit. It will be demonstrated later that the desired phase shift covering full $2\pi$ range can be achieved by adjusting a single parameter of $t$, which guarantees a totally planar metasurface. The parameters of $a_x$ and $a_y$ are fixed to be $\lambda_0/8$ with $\lambda_0 = 0.1\text{m}$ to ensure the subwavelength feature of the resulting device. The width of these identical plates $w$ (gray regions in Fig. 1(c)) is chosen as 1 mm for facilitating the sample fabrication. Figure 1(d) illustrates the phase change as a function of $t$ for 4 kinds of labyrinthine units. Actually, by appropriately choosing the value of $t$, the case $(m,n) = (3,2)$ is sufficient for providing a phase change ranging from 0 to $2\pi$ with a step of $\pi/4$. In the experimental demonstrations, labyrinthine units with less zigzagged structures are employed for facilitating the fabrication of samples as well as minimizing the loss viscosity effect. According to the phase shift shown in Fig. 1(d), 8 units could be chosen covering $2\pi$ range with a step $\pi/4$ and the values of $t$ for each unit are labeled with 8 dots. Then the width of the labyrinthine channel can be expressed as $d = a_x/(m+n) - w$, and the length of the plates is $l = a_y - 2t - d$ for these 8 types of unit. Note that by employing common anisotropic metamaterials, it is still possible to realize the similar results of wavefront manipulation approximately. In such cases, however, the relatively low refractive indexes significantly limit the amount of phase shift that can be provided within a certain distance [34, 36]. As a result, the total thickness of the whole system based on these materials must be considerably large for realizing the desired phase shift. The labyrinthine units used here take advantage of high refractive index [16, 37-40] and the subsequently small thickness ($\lambda_0/8$), which enables the design of an acoustical metasurface.

**C. Sample fabrication and experimental setup**

Figure 1(e) schematically demonstrates the experimental setup. The metasurfaces are fabricated with thermoplastics via 3D printing (Stratasys Dimension Elite, 0.08mm in precision) to meet the requirement of the theoretical design. The sample are placed between two paralleled plexiglass plates (1.2m × 1.5m), which forms the 2D experimental environment, and the setup is schematically shown in Fig. 1(e). The distance of the experimental system in $z$ direction is 2.1 cm as same as the thickness of the metasurfaces so that the sample can contact with the two plates close together. Wedge-shaped sound-absorbing foam is installed at the boundary of the plexiglass plates to mimic an anechoic environment. A loudspeaker (30mm in diameter) is mounted in a box baffle and placed outside but very near to the upper plate one meter away from the metasurfaces. It emits the monochromatic wave as the acoustic source and propagates into the 2D experimental environment between the paralleled plexiglass plates through a small hole (20mm in diameter) in the upper plate. For each measurement, two 1/4-inch microphones (Brüel & Kjær type-4961) are situated at designated positions to sense the local pressure: one is mounted at a fixed position to detect the pressure as signal 1, and the other is moveable to scan the pressure field as signal 2. By using the software PULSE Labshop, the cross-spectrum of the two signals gathered by the two microphones is obtained, for which signal 1 works as a reference and signal 2 as an input signal. The pressure field is retrieved by analyzing the cross-spectrum and recording the magnitude and phase at different spatial position within the measured region. Then the reflected pressure field is obtained by subtracting two pressure fields with and without the metasurfaces.

## III.   RESULTS AND DISCUSSIONS

### A. The generalized Snell's law of reflection and anomalous reflection

The full $2\pi$ range phase shift offered by the metasurface comprising a monolayer of metamaterial enables revisiting the Snell's law. Here we will combine theory and experiment to demonstrate that a properly designed metasurface can support anomalous reflection for incident acoustic waves governed by the generalized Snell's law. The schematics for the derivation of the angle of reflection are shown in Fig. 2(a), where $\theta_r$ represents the reflection angle. For supporting the anomalous reflection, the phase gradient along $y$ direction should satisfy $d\phi(y)/dy = -k_0 \sin\theta_r$. Here, the reflection angle is set to be $\pi/6$. In this case, since the incident waves are radiated from a loudspeaker located at $(x', y') = (1, 0)$, there exists an extra phase change $\phi_{extra} = -k_0\left(\sqrt{y^2+1}-1\right)$ that needs to be compensated by the metasurface as well. Therefore, the desired phase profile yielded by the metasurface should be $\phi(y) = -k_0(y+0.3)\sin(\pi/6) - k_0\left(\sqrt{y^2+1}-1\right)$ as shown in Fig 2(b). Such a profile can then be yielded by constructing the acoustic metasurface with appropriately designed labyrinthine units. The pressure field distribution predicted analytically by Eq. (2) is plotted in Fig. 2(c), where the black box indicates the region within which the measurement is performed. The results of the experiments, along with that of the numerical simulations for the metasurface composed by labyrinthine units, are shown in Figs. 2(d) and 2(c) respectively. Excellent agreement is observed between the numerical and the experimental results with the black arrows referring to the theoretical angle of the reflection. They both demonstrate that by carefully designing the geometric parameters, the acoustic metasurface can bend incident wave to the desired angle, following the generalized Snell's law with an additional parallel wave-vector provided by the phase gradient of the metasurface and, furthermore, can even be used to realize the conversion from propagating waves to surface waves[1, 4]. The generalization of the law of reflection by acoustic metasurface is applicable to a large variety of interfaces with subwavelength engineered structures, offer novel opportunity to control over acoustic waves versatilely.

## B. Ultrathin Planar acoustic lens

Figure 3(a) illustrates the schematic of the design of a metasurface-based lens. For obtaining a focal spot in the axis located at $(x, y) = (f_x, 0) = (0.3, 0)$, the equiphase surface should be the gray line, indicating that the phase profile should be expressed as $\phi(y) = -k_0 \left( \sqrt{y^2 + f_x^2} - f_x \right)$, and the phase profile provided by the metasurface should be $\phi(y) = -k_0 \left( \sqrt{y^2 + 0.9} - 0.3 \right) - k_0 \left( \sqrt{y^2 + 1} - 1 \right)$, as shown in Fig. 3(b). The pressure field distribution predicted analytically by Eq. (2) is shown in Fig. 3(c), showing the effectiveness of such a design. The experimental results, which are in excellent agreement with the numerical simulation, demonstrate that the metasurface can be tailored to mimic an ultrathin planar lens that has a thickness only $\lambda_0/8$ of the incident wavelength yet is capable of effectively steering the convergence of acoustic waves. For quantifying the performance of the lens, acoustic intensity distributions along the axis and the transverse cross-section through the focal point are measured and shown in Figs. 3(c) and 3(d), which agree almost perfect with the theoretical and numerical predictions. Furthermore, due to the full control of the phase profile, the metasurface can be conveniently reconfigured to become an acoustic lens with arbitrary focal point. The corresponding results for a typical example of decentered lens are illustrated in Fig. 4 with focal point at $(x, y) = (f_x, f_y) = (0.3, 0.1)$, in which perfect match between the theoretical and measured results can be observed as well. The capability of focusing low-frequency sound using subwavelength-scale metasurface that manipulates the phase of propagating wave may pave the way to diverse applications such as acoustic imaging and caustic engineering that needs design of the shape of a focused trajectory of sound. The focusing effect can also be realized by metamaterial-based lenses, however, such lenses may suffer from the limitation in refractive index, resulting in considerably large size [36]. In comparison, the high refractive index provided by the proposed acoustic metasurface enables full control over the phase of wave with a lens thickness as small as $\lambda_0/8$, which offers greater freedom to design the focal point.

### C. Self-bending beams with arbitrary trajectory

The accelerating optical beams, such as the Airy beams [41, 42], the non-paraxial Mathieu and Webber beams [43], etc., have recently been experimentally demonstrated and have significant impact in the optical fields. In order to show the potential of the acoustic metasurface to fully control the acoustic waves, we demonstrate both theoretically and experimentally the generation of self-accelerating/bending acoustic beam that propagates along an arbitrary caustic trajectory [44, 45]. It is worth emphasizing that the proposed scheme can be used to design metasurfaces with the ability to generate paraxial acoustic beams, represented by Airy beams, as well as the non-paraxial ones. For facilitating the experimental demonstration, however, here we have only investigated the design and fabrication of metasurface capable of generating non-paraxial beam, instead of the case of paraxial beams which are less general but much harder to measure. Figure 5(a) illustrates the schematic of the design of a metasurface which is in fact a phase mask to convert an incident wave to a non-paraxial beam based on the caustic theory [44]. The relationship between the phase profile $\phi(y)$ and the reflected angle $\theta_r$ in $y$ direction could be expressed as $d\phi(y)/dy = -k_0 \sin\theta_r$. To realize the arbitrary trajectory $y = f(x)$, the phase profile can be expressed as $d\phi(y)/dy = -k_0 \left[ f'(x)/\sqrt{1+(f'(x))^2} \right]$ with $f'(x) = \tan\theta_r$ being the slope of the trajectory. For a particular half circle trajectory $f(x) = \sqrt{r^2 - (x-r)^2}$ with center at $(x, y) = (r, 0)$, the desired profile could be expressed as $\phi(y) = -k_0 \left( y - 2r\sqrt{y/r} \right)$. In the simulations and experiments, the length of the metasurface is $8\lambda_0$ and extends from $y=0$ to $y=8\lambda_0$ and the point source locates at $(x, y) = (1, 0.4)$. Then the desired phase profile yielded by metasurface should be $\phi(y) = -k_0 \left[ \left( y - 0.7\sqrt{y/0.35} \right) + \left( \sqrt{(y-0.4)^2 + 1} - 1 \right) \right]$ with $r = 0.35$ m, as shown in Fig. 5(b). The pressure field distribution analytically predicted by Eq. (3) is shown in Fig. 5(c). The numerical simulations of the metasurface composed by appropriately choosing the labyrinthine unit are shown in Fig. 5(d), and the corresponding experimental results are shown in Fig. 5(e). Due to the fine resolution along $y$ direction in our simulations and experiments (viz. $\lambda_0/8$), good agreement is achieved between the results of numerical simulation and experimental measurement. Trivial spatial aliasing effect [46], which accounts for the slight difference between the analytical predictions and the numerical and experimental results, can also observed due to the inevitable discretization and the quick oscillation of the desired phase profile along $y$ direction, as shown in Fig. 5(b).

    We anticipate the acoustic metasurface capable of yielding arbitrary acoustic beams to provide remarkable flexibility and open new exciting possibilities in a variety of acoustic-based applications ranging from ultrasound imaging to acoustic device design. For instance, these beams show promise to realize the trapping and guiding of microparticles along the arbitrary trajectories, or they may enable circumventing an acoustic obstacle by designing a bypassing caustic.

## IV. CONCLUSION

In conclusion, we have designed and experimentally realized acoustic metasurfaces that can fully control the reflected waves in direction and shape of wavefront. A general scheme of designing an acoustic metasurface is developed on the basis of labyrinthine units, which can give analytical description to the phase shifts. By suitably designing the geometry of the metamaterial units, the metasurface with subwavelength can reflect acoustic waves in an unusual yet controllable manner. As particular examples, three kinds of distinct properties of acoustic wave manipulation have been demonstrated both theoretically and experimentally, yielding anomalous reflection governed by the generalized Snell's law, mimicking ultrathin planar lens and, remarkably, generating non-diffractive beams along arbitrary trajectory. The results reveal the great flexibility in the design of acoustic beams enabled by the free manipulation of propagating phase with metasurface. The realization of the acoustic metasurface with the capability of controlling acoustic waves arbitrarily should open the avenue for designing versatile compact acoustic components and have potential application in a large variety of fields in which special manipulation of acoustic flux is required, such as ultrasound imaging and caustic engineering.

## ACKNOWLEDEMENTS

We acknowledge helpful discussions with Prof. Tao Li. This work was supported by the National Basic Research Program of China (973 Program) (Grant Nos. 2010CB327803 and 2012CB921504), National Natural Science Foundation of China (Grant Nos. 11174138, 11174139, 11222442, 81127901, and 11274168), NCET-12-0254, and A Project Funded by the Priority Academic Program

**References**

[1] N. Yu, P. Genevet, M. A. Kats, F. Aieta, J. P. Tetienne, F. Capasso, and Z. Gaburro, Light propagation with phase discontinuities: Generalized laws of reflection and refraction, Science **334**, 333 (2011).

[2] F. Aieta, P. Genevet, M. A. Kats, N. Yu, R. Blanchard, Z. Gaburro, and F. Capasso, Aberration-free ultrathin flat lenses and axicons at telecom wavelengths based on plasmonic metasurfaces, Nano Lett. **12**, 4932 (2012).

[3] X. Ni, N. K. Emani, A. V. Kildishev, A. Boltasseva, and V. M. Shalaev, Broadband light bending with plasmonic nanoantennas, Science **335**, 427 (2012).

[4] S. Sun, Q. He, S. Xiao, Q. Xu, X. Li, and L. Zhou, Gradient-index meta-surfaces as a bridge linking propagating waves and surface waves, Nat. Mater. **11**, 426 (2012).

[5] F. Monticone, N. M. Estakhri, and A. Alu, Full control of nanoscale optical transmission with a composite metascreen, Phys. Rev. Lett. **110**, 203903 (2013).

[6] H. Wakatsuchi, S. Kim, J. J. Rushton, and D. F. Sievenpiper, Waveform-dependent absorbing metasurfaces, Phys. Rev. Lett. **111**, 245501 (2013).

[7] Y. Mazor, and B. Z. Steinberg, Metaweaves: Sector-way nonreciprocal metasurfaces, Phys. Rev. Lett. **112**, 153901 (2014).

[8] P. M. Morse, and K. U. Ingard, *Theoretical Acoustics* (Princeton University Press, 1986).

[9] Z. Liu, X. Zhang, Y. Mao, Y. Y. Zhu, Z. Yang, C. T. Chan, and P. Sheng, Locally resonant sonic materials, Science **289**, 1734 (2000).

[10] J. Li, and C. T. Chan, Double-negative acoustic metamaterial, Phys. Rev. E **70**, 055602 (2004).

[11] N. Fang, D. Xi, J. Xu, M. Ambati, W. Srituravanich, C. Sun, and X. Zhang, Ultrasonic metamaterials with negative modulus, Nat. Mater. **5**, 452 (2006).

[12] Y. Ding, Z. Liu, C. Qiu, and J. Shi, Metamaterial with simultaneously negative bulk modulus and mass density, Phys. Rev. Lett. **99**, 093904 (2007).

[13] Z. Yang, J. Mei, M. Yang, N. H. Chan, and P. Sheng, Membrane-type acoustic metamaterial with negative dynamic mass, Phys. Rev. Lett. **101**, 204301 (2008).

[14] S. H. Lee, C. M. Park, Y. M. Seo, Z. G. Wang, and C. K. Kim, Composite acoustic medium with simultaneously negative density and modulus, Phys. Rev. Lett. **104**, 054301 (2010).

[15] J. Christensen, and F. J. Garcia de Abajo, Anisotropic metamaterials for full control of acoustic waves, Phys. Rev. Lett. **108**, 124301 (2012).

[16] Z. Liang, and J. Li, Extreme acoustic metamaterial by coiling up space, Phys. Rev. Lett. **108**, 114301 (2012).

[17] Y. Xie, B. I. Popa, L. Zigoneanu, and S. A. Cummer, Measurement of a broadband negative index with space-coiling acoustic metamaterials, Phys. Rev. Lett. **110**, 175501 (2013).

[18] M. Yang, G. Ma, Z. Yang, and P. Sheng, Coupled membranes with doubly negative mass density and bulk modulus, Phys. Rev. Lett. **110**, 134301 (2013).

[19] Z. Liang, T. Feng, S. Lok, F. Liu, K. B. Ng, C. H. Chan, J. Wang, S. Han, S. Lee, and J. Li, Space-coiling metamaterials with double negativity and conical dispersion, Sci. Rep. **3**, 1614 (2013).

[20] J. Li, L. Fok, X. Yin, G. Bartal, and X. Zhang, Experimental demonstration of an acoustic magnifying hyperlens, Nat. Mater. **8**, 931 (2009).

[21] S. Zhang, L. Yin, and N. Fang, Focusing ultrasound with an acoustic metamaterial network, Phys. Rev. Lett. **102**, 194301 (2009).

[22] J. Zhu, J. Christensen, J. Jung, L. Martin-Moreno, X. Yin, L. Fok, X. Zhang, and F. J. Garcia-Vidal, A holey-structured metamaterial for acoustic deep-subwavelength

imaging, Nat. Phys. **7**, 52 (2011).

[23] B. Liang, B. Yuan, and J. C. Cheng, Acoustic diode: Rectification of acoustic energy flux in one-dimensional systems, Phys. Rev. Lett. **103**, 104301 (2009).

[24] B. Liang, X. S. Guo, J. Tu, D. Zhang, and J. C. Cheng, An acoustic rectifier, Nat. Mater. **9**, 989 (2010).

[25] X. F. Li, X. Ni, L. A. Feng, M. H. Lu, C. He, and Y. F. Chen, Tunable unidirectional sound propagation through a sonic-crystal-based acoustic diode, Phys. Rev. Lett. **106**, 084301 (2011).

[26] Y. Li, B. Liang, Z. M. Gu, X. Y. Zou, and J. C. Cheng, Unidirectional acoustic transmission through a prism with near-zero refractive index, Appl. Phys. Lett. **103**, 053505 (2013).

[27] R. Fleury, D. L. Sounas, C. F. Sieck, M. R. Haberman, and A. Alu, Sound isolation and giant linear nonreciprocity in a compact acoustic circulator, Science **343**, 516 (2014).

[28] B. I. Popa, L. Zigoneanu, and S. A. Cummer, Experimental acoustic ground cloak in air, Phys. Rev. Lett. **106**, 253901 (2011).

[29] S. Zhang, C. Xia, and N. Fang, Broadband acoustic cloak for ultrasound waves, Phys. Rev. Lett. **106**, 024301 (2011).

[30] L. Zigoneanu, B. I. Popa, and S. A. Cummer, Three-dimensional broadband omnidirectional acoustic ground cloak, Nat. Mater. **13**, 352 (2014).

[31] J. Mei, G. Ma, M. Yang, Z. Yang, W. Wen, and P. Sheng, Dark acoustic metamaterials as super absorbers for low-frequency sound, Nat. Commun. **3**, 756 (2012).

[32] G. Ma, M. Yang, S. Xiao, Z. Yang, and P. Sheng, Acoustic metasurface with hybrid resonances, Nat. Mater. **13**, 873 (2014).

[33] Y. Li, B. Liang, Z. M. Gu, X. Y. Zou, and J. C. Cheng, Reflected wavefront manipulation based on ultrathin planar acoustic metasurfaces, Sci. Rep. **3**, 2546 (2013).

[34] J. Zhao, B. Li, Z. Chen, and C. W. Qiu, Manipulating acoustic wavefront by inhomogeneous impedance and steerable extraordinary reflection, Sci. Rep. **3**, 2537 (2013).

[35] Y. Xie, A. Konneker, B.-I. Popa, and S. A. Cummer, Tapered labyrinthine acoustic metamaterials for broadband impedance matching, Appl. Phys. Lett. **103**, 201906 (2013).

[36] F. Cervera, L. Sanchis, J. V. Sanchez-Perez, R. Martinez-Sala, C. Rubio, F. Meseguer, C. Lopez, D. Caballero, and J. Sanchez-Dehesa, Refractive acoustic devices for airborne sound, Phys. Rev. Lett. **88**, 023902 (2002).

[37] Y. Li, B. Liang, X. Tao, X. F. Zhu, X. Y. Zou, and J. C. Cheng, Acoustic focusing by coiling up space, Appl. Phys. Lett. **101**, 233508 (2012).

[38] Y. Li, B. Liang, X. Y. Zou, and J. C. Cheng, Extraordinary acoustic transmission through ultrathin acoustic metamaterials by coiling up space, Appl. Phys. Lett. **103**, 063509 (2013).

[39] T. Frenzel, J. D. Brehm, T. Buckmann, R. Schittny, M. Kadic, and M. Wegener, Three-dimensional labyrinthine acoustic metamaterials, Appl. Phys. Lett. **103**, 061907 (2013).

[40] Y. Li, G. K. Yu, B. Liang, X. Y. Zou, G. Y. Li, S. Cheng, and J. C. Cheng, Three-dimensional ultrathin planar lenses by acoustic metamaterials, Sci. Rep. **4**, 6830 (2014).

[41] G. A. Siviloglou, and D. N. Christodoulides, Accelerating finite energy airy beams, Opt. Lett. **32**, 979 (2007).

[42] L. Li, T. Li, S. M. Wang, C. Zhang, and S. N. Zhu, Plasmonic airy beam generated by in-plane diffraction, Phys. Rev. Lett. **107**, 126804 (2011).

[43] P. Zhang, Y. Hu, T. Li, D. Cannan, X. Yin, R. Morandotti, Z. Chen, and X. Zhang, Nonparaxial mathieu and weber accelerating beams, Phys. Rev. Lett. **109**, 193901 (2012).

[44] E. Greenfield, M. Segev, W. Walasik, and O. Raz, Accelerating light beams along arbitrary convex trajectories, Phys. Rev. Lett. **106**, 213902 (2011).

[45] P. Zhang, T. Li, J. Zhu, X. Zhu, S. Yang, Y. Wang, X. Yin, and X. Zhang, Generation of acoustic self-bending and bottle beams by phase engineering, Nat. Commun. **5**, 4316 (2014).

[46] D. H. Johnson, and D. E. Dudgeon, *Array Singal Processing: Concepts and Techniques* (Prentice Hall, 1993).

**Figure Captions:**

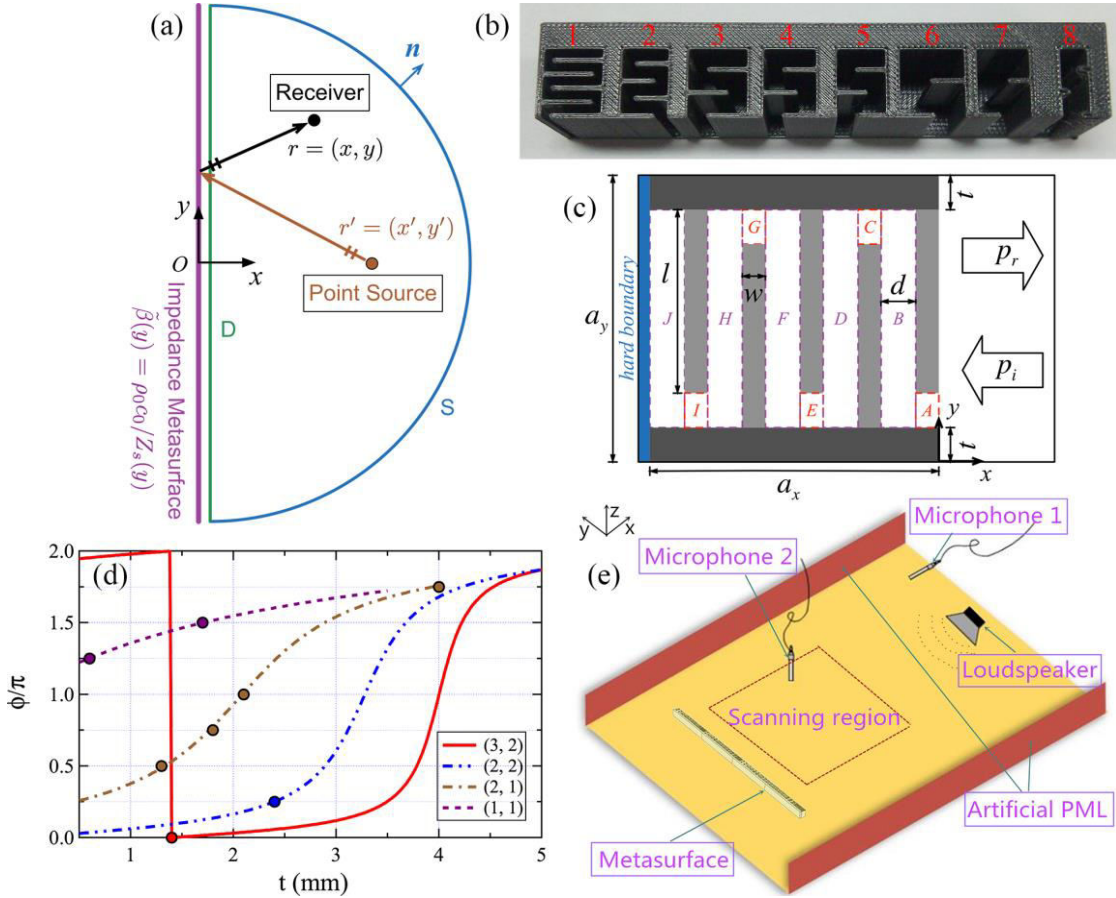

Fig. 1. (a) The sketch map for deriving the reflected pressure field with a point source. The proposed metasurface is characterized by acoustic admittance $\tilde{\beta}(y)$. (b) Photo of proposed 8 labyrinthine units labeled with "1" to "8", which could provide phase shift ranging from 0 to $2\pi$ with a step of $\pi/4$. (c) The details of a specific labyrinthine unit (width $a_x$ and height $a_y$), which are composed by several cross arranged identical plates (width $w$ and length $l$) in which way to form a labyrinthine channel with width $d$. Another two plates (thickness $t$) are used to seal the channels in $y$ direction. (d) The phase shifts induced by 4 different kinds of labyrinthine units as a function of $t$. The $(m,n)$ case revealing that there are $m$ identical plates in the upper boundary and $n$ identical plates in the down boundary. 8 dots refer to the chosen 8

units shown in (b). (e) The experimental setup for scanning the pressure fields.

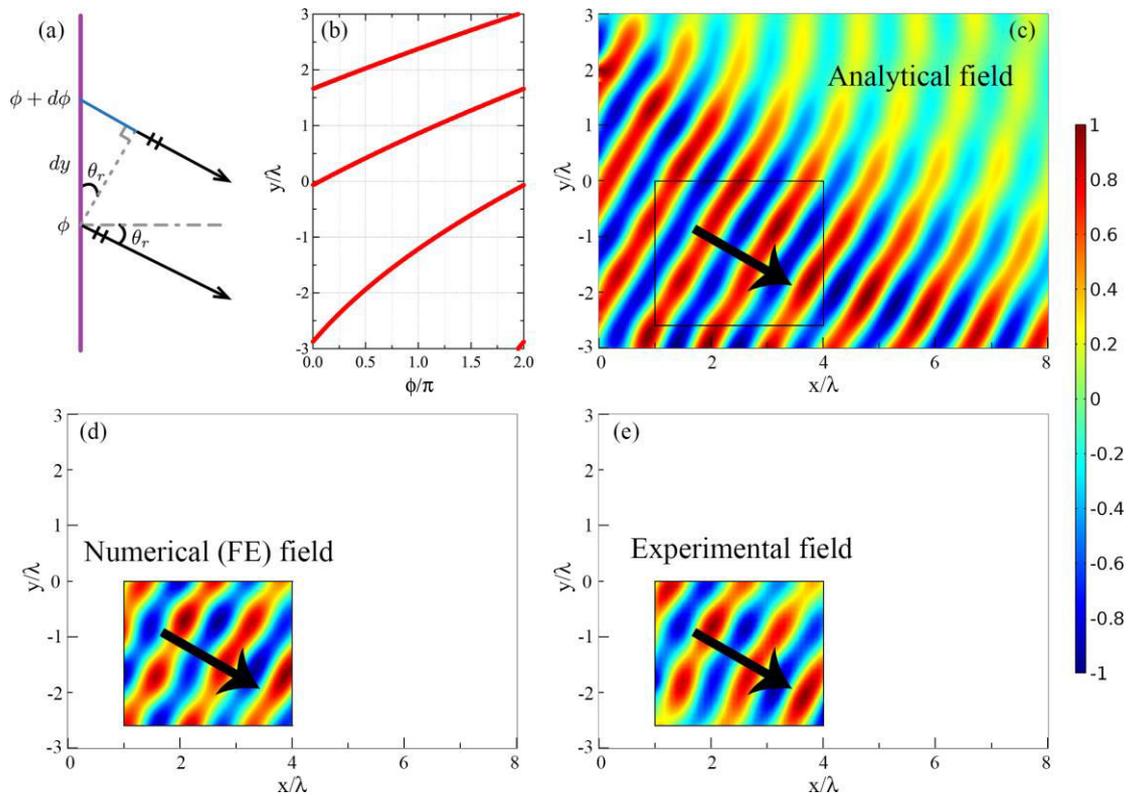

Fig. 2. (a) The schematics for the derivation of the angle of reflection. The length of blue line refers to the difference of the acoustic path distance provided by the metasurface in $y$ direction. (b) The desired continuous phase profile that needs to be yielded by the metasurfaces with length $2L = 6\lambda$. (c) The pressure field distribution predicted by the Green's function theory. The black box indicates the region within which the measurement is performed. (d) The numerical simulation of the metasurface composed labyrinthine units and (e) the experimental results of the pressure field distribution. Black arrows in the (c-e) refer to the theoretical angle of the reflection.

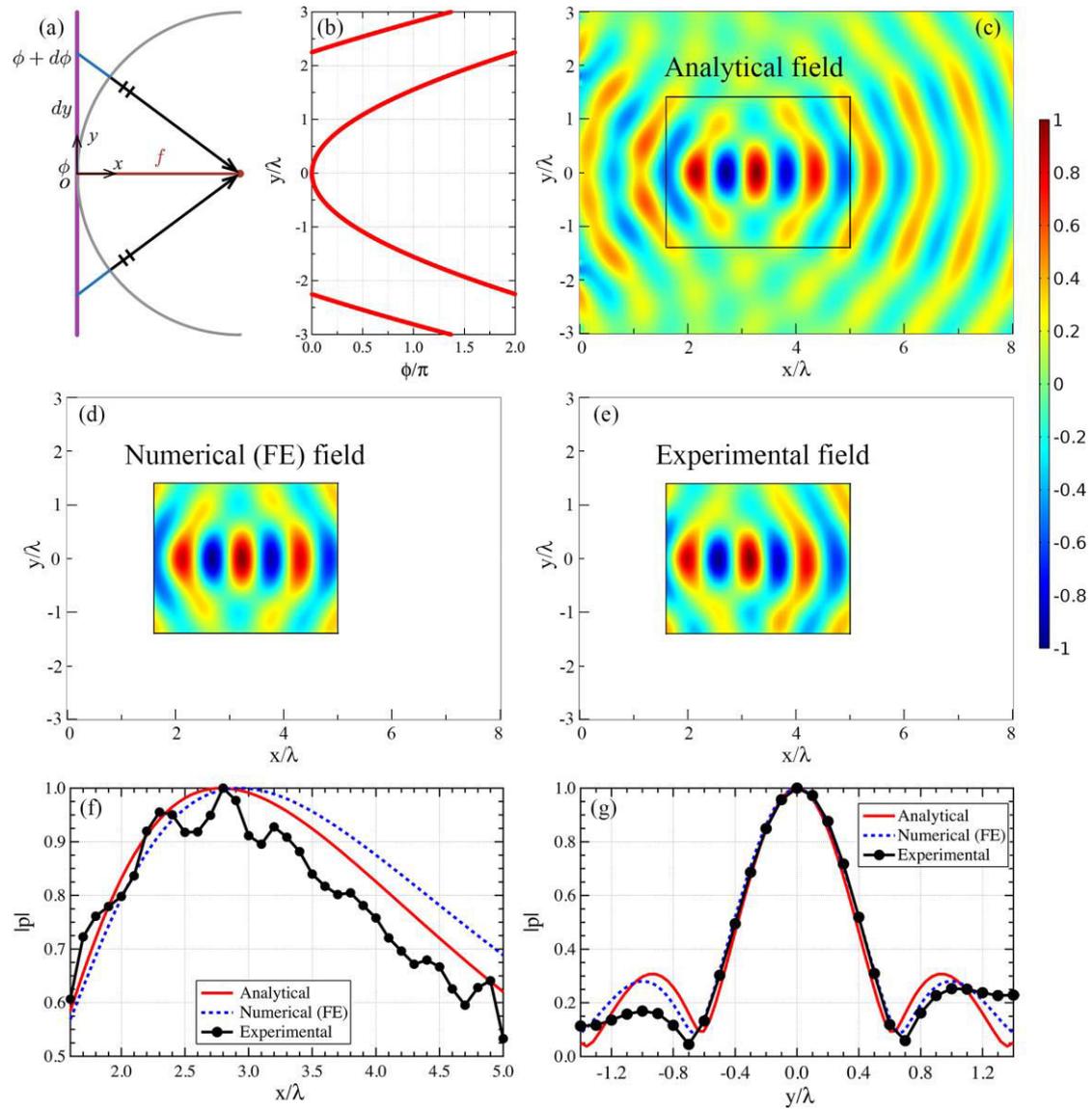

Fig. 3. (a) The schematics for the derivation of the acoustic focusing with focus length $f$. The length of blue line refers to the difference in the acoustic path distance provided by the metasurface along $y$ direction. (b) The desired continuous phase profile that needs to be yielded by the metasurfaces with length $2L=6\lambda$. (c) The pressure field $p(x,y)$ distribution predicted by the Green function theory. The black box indicates the region within which the measurement is performed. (d) The numerical simulation of metasurface composed labyrinthine units and (e) the experimental results of the pressure field distribution. The amplitude of pressure $|p|$

profile (f) along the axis and (g) transverse the axis through the focal points.

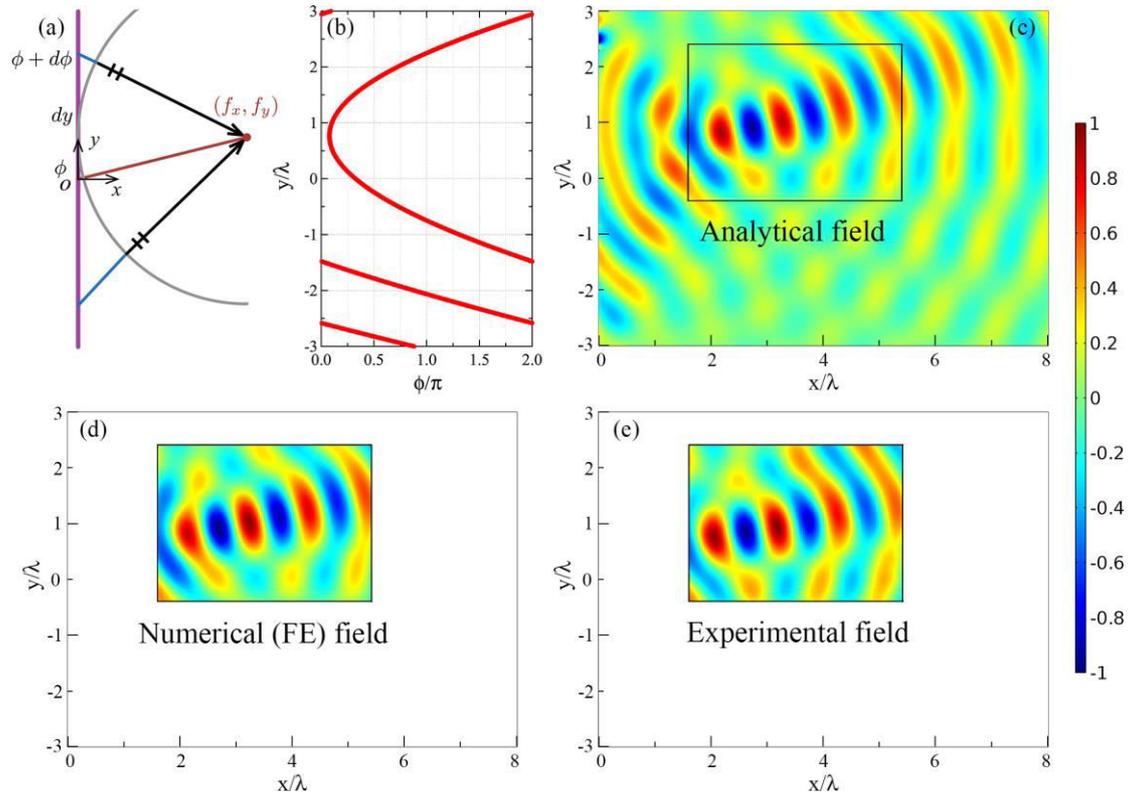

Fig. 4. (a) The schematics for the derivation of the acoustic focusing with focus point locating at $(f_x, f_y)$. The length of blue line refers to the acoustic path distance difference provided by the metasurface. (b) The desired continuous phase profile that needs to be yielded by the metasurfaces with length $2L = 6\lambda$. (c) The pressure field $p(x, y)$ distribution predicted by the Green's function theory. The black box indicates the region within which the measurement is performed. (d) The numerical simulation of metasurface composed labyrinthine units and (e) the experimental results of the pressure field distribution.

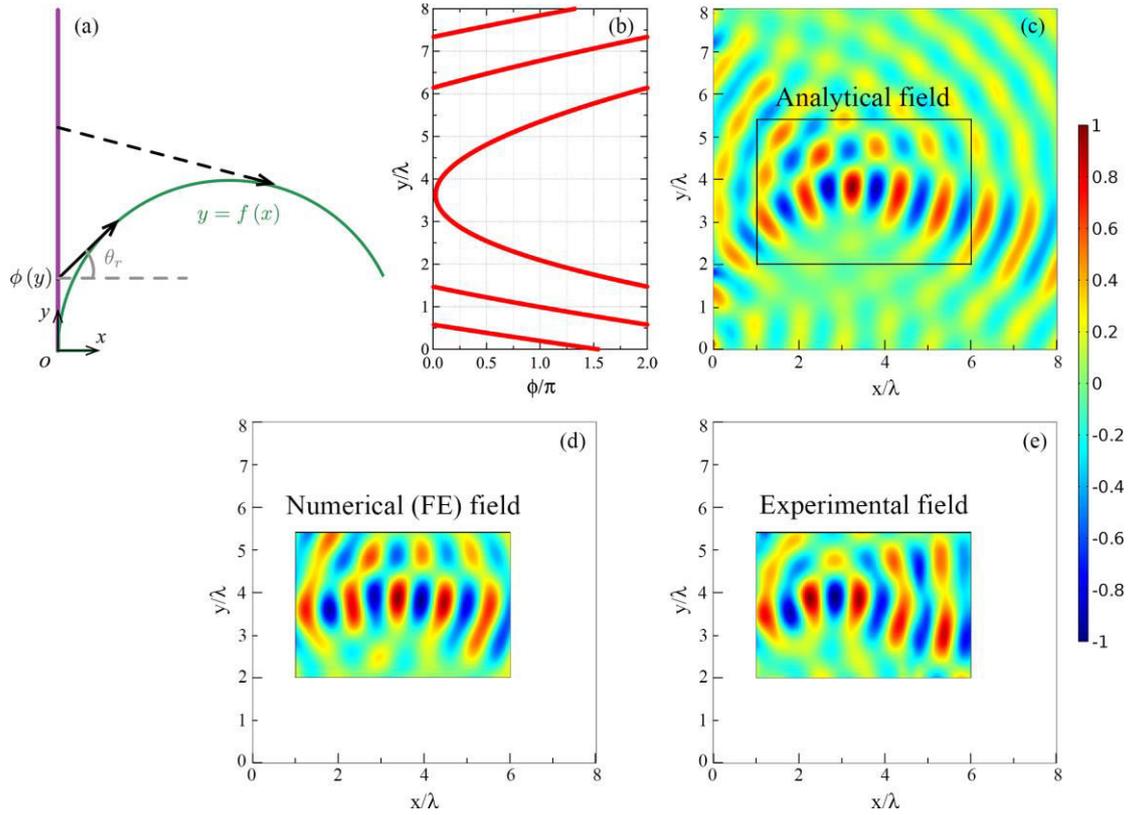

Fig. 5. (a) The schematics for the derivation of the phase profile for the arbitrary trajectory. (b) The desired continuous phase profile that needs to be yielded by the metasurfaces with length $2L = 8\lambda$. (c) The pressure field $p(x,y)$ distribution predicted by the Green function theory. The black box indicates the region within which the measurement is performed. (d) The numerical simulation of the metasurface composed by labyrinthine units and (e) the experimental results of the pressure field distribution.